\documentclass[12pt,twoside]{article}
\usepackage[mathscr]{eucal}
\usepackage{amsmath,amsfonts,amssymb,amsthm}
\usepackage{dutchcal}

\usepackage{times}
\usepackage{amsmath}
\usepackage{amsfonts}
\usepackage{epstopdf}
\usepackage{hyperref}

\advance\textheight\topskip
\renewcommand{\tilde}{\widetilde}
\renewcommand{\hat}{\widehat}

\renewcommand{\d}{\partial}

\newcommand*\xbar[1]{%
  \hbox{%
    \vbox{%
      \hrule height 0.5pt 
      \kern0.3ex
      \hbox{%
        \kern-0.0em
        \ensuremath{#1}%
        \kern-0.0em
      }%
    }%
  }%
} 
\numberwithin{equation}{section} \makeatletter
\@addtoreset{equation}{section}

\begin{document}

\def\mytitle{The BMS4 algebra at spatial infinity.}

\pagestyle{myheadings} \markboth{\textsc{\small Troessaert}}{%
  \textsc{\small The BMS algebra at spatial infinity.}} \addtolength{\headsep}{4pt}

\begin{centering}

  \vspace{1cm}

  \textbf{\Large{\mytitle}}



  \vspace{1.5cm}

  {\large C\'edric Troessaert$^a$}

\vspace{.5cm}

\begin{minipage}{.9\textwidth}\small \it \begin{center}
   Max Planck Institute for Gravitational Physics\\ Am M\"uhlenberg 1,
   Golm, Germany \\ ctroessaert@aei.mpg.de \end{center}
\end{minipage}

\end{centering}

\vspace{1cm}

\begin{center}
  \begin{minipage}{.9\textwidth}
    \textsc{Abstract}. 
    We show how a global BMS4 algebra
	  appears as part of the asymptotic symmetry algebra at spatial infinity.
	  Using linearised theory, we then show that this global BMS4 algebra
	  is the one introduced by Strominger as a symmetry
	  of the S-matrix.
  \end{minipage}
\end{center}

\vfill

\noindent
\mbox{}

\thispagestyle{empty}
\newpage


\section{Introduction}
\label{sec:introduction}

In the last few years, following the work of Strominger and collaborators
\cite{Strominger2013,Strominger2014,He2015}, 
a new understanding of infrared divergences in
scattering processes has appeared. It was shown that soft
theorems are related to Ward identities derived from conserved charges
associated to
asymptotic symmetries at null infinity (see for instance
\cite{Hyun2014,Adamo2014,Kapec2014,Campiglia2015,Campiglia2015a,Kapec2015,Avery2015,Campiglia2015b,Avery2015a,Strominger2017}).
One of the main examples where this relation
appeared is between the BMS4 algebra of asymptotically flat space-times at
null infinity and scattering processes of gravitons. It was shown that Weinberg's soft gravitons theorem
can be seen as the Ward identity derived from super-translation charges.

More recently, Hawking, Perry and Strominger
 reconsidered the information loss problem in the evaporation of Black-holes
\cite{Hawking2016a,Hawking2016}. The existence of the infinite set of
conserved charges associated to BMS4 means that part of the information about the system is
retained in the form of global/soft gravitons. This will imply correlations
between Hawking radiation produced by the black-hole and the status of the 
system before the collapse.

A key point of the results described above is the existence of a global BMS4 algebra and its
associated conserved charges during an evolution process.
This existence derives from a set of junction conditions at
spatial infinity between various fields defined at past and future null
infinity. While these hypothesises are well motivated and are equivalent to the soft
graviton theorem \cite{He2015,Kapec2014}, their existence is
surprising considering the non-differentiability of spatial infinity.
These facts mean that both the symmetries and their associated charges are only
defined at null infinity. They are properties of the initial and final states
but we, in general, don't have a good understanding of them at finite times.

Recently, a 
description of the asymptotic symmetry algebra of electromagnetism
has been done at spatial infinity by Campiglia and Eyheralde
\cite{Campiglia2017}. Their
description provides a bridge between both asymptotic regimes and proves the
junction condition in the context of electromagnetism. 

In the case of gravity, it has been shown in \cite{Herberthson1992}
that the junction condition for the mass aspect between future and past null infinity is a 
consequence of the structure of spatial infinity. 
This result hints at the fact that there should also exist a description of
spatial infinity for which the global BMS4 symmetry introduced by Strominger
appears naturally as an asymptotic symmetry algebra.

A similar analysis has been done in 3 dimensions by Compere and
Fiorucci \cite{Compere2017}. Their results are the 3 dimensional equivalent to
the ones we report here.

\vspace{5mm}

In this paper, we consider the set of asymptotic conditions at spatial infinity introduced by
Comp\`ere and Dehouck in \cite{Compere2011}. These asymptotic conditions are a
generalisation of the results obtained  in
\cite{Mann2006,Mann2006a,Mann2008} for the holographic renormalization of
asymptotically flat space-times. Our main result is that a sub-algebra of the
associated asymptotic symmetries is a global non extended BMS4. Using the linearised theory
around flat space, we then show that, when
restricting the analysis to asymptotically flat space-times at null infinity,
this global non extended BMS4 algebra defined at spatial infinity is identical to
the one obtained by Strominger in his original analysis \cite{Strominger2014}. We also show that the associated conserved
charges defined at spatial infinity reproduce the BMS4 charges at null infinity.

\vspace{5mm}

This paper is organized as follows. In section \ref{sec:spatialinfinity}, we introduce the
asymptotic conditions of Comp\`ere and Dehouck. In section \ref{sec:bms4alg}, we study the
asymptotic symmetry algebra and show that it contains BMS4. In the last
section, we make the link with null infinity using the linearised theory and a
description of space-like infinity introduced by Friedrich \cite{Friedrich1998}.

\section{Spatial infinity}
\label{sec:spatialinfinity}

Our analysis uses the description of spatial infinity introduced by
Comp\`ere and Dehouck in \cite{Compere2011}. It corresponds to a hyperbolic
description of spatial infinity with relaxed asymptotics.
The metrics they consider are a generalisation of the Beig-Schmidt ansatz \cite{Beig1982}:
\begin{multline}
	\label{eq:metricI}
	g_{\mu\nu} dx^\mu dx^\nu = \left(1 + \frac{2\sigma}{\eta} +
	\frac{\sigma^2}{\eta^2} + o(\eta^{-2})\right)
	d\eta^2 + o(\eta^{-1}) d\eta dx^b\\
	+\left(\eta^2 h_{ab}^{(0)} + \eta (k_{ab} - 2 \sigma h_{ab}^{(0)}) + \log \eta
	i_{ab} + h_{ab}^{(2)} + o(\eta^0)\right) dx^adx^b,
\end{multline}
where the asymptotic gravitational fields, $\sigma, k_{ab}, i_{ab}$ and
$h_{ab}^{(n)}$ depend on $x^a$ only. Spatial infinity is reached in the limit
$\eta \to \infty$ with the corresponding boundary metric $h^{(0)}_{ab}$ being the 
metric on the unit hyperboloid. We
will write it as:
\begin{equation}
	x^a = (s, x^A), \qquad h^{(0)}_{ab} dx^a dx^b = \frac{-1}{(1-s^2)^2}
	ds^2 + \frac{1}{1-s^2} \gamma_{AB} dx^Adx^B,
\end{equation}
with $x^A$ and $\gamma_{AB}$ respectively coordinates and the unit metric on the
2-sphere. We will use $h_{ab}^{(0)}$ and its inverse to lower and raise
indices of the asymptotic quantities defined on the hyperboloid and denote the associated
covariant derivative by $\mathcal D_a$.

The diffeomorphisms preserving this
form of the metric contain the transformations generated by 
\begin{flalign}
	\label{eq:asymptoticsimI}
	\xi^\eta &= - \omega +\frac{1}{\eta}\omega^{(1)}+ o(\eta^{-1}), \qquad
	\omega^{(1)} = \omega \sigma - \mathcal D^a \omega \d_a \sigma,\\
	\label{eq:asymptoticsimII}
	\xi^a &= \mathcal Y^a -\frac{1}{\eta} \mathcal D^a
	\omega +\frac{1}{2\eta^2} \left( k^{ab} \d_b \omega-4\sigma \mathcal D^a \omega
	+ \mathcal D^a \omega^{(1)}\right)+ o(\eta^{-2}),
\end{flalign}
where $\mathcal Y = \mathcal Y^a(x^b) \d_a$ is a killing vector of
$h_{ab}^{(0)}$ and $\omega(x^a)$ is an arbitrary function on the hyperboloid.
The transformation parametrised by $\omega$ form the abelian sub-algebra
of SPI super-translations \cite{Ashtekar1978} while the vectors $\mathcal Y$
parametrise the Lorentz sub-algebra.
On top of these, the asymptotic form of the metric given in \eqref{eq:metricI} is
also invariant under transformations called logarithmic translations.
They are parametrised by a function $H(x^a)$ on the hyperboloid
satisfying
\begin{equation}
	\label{eq:eqlogtrans}
	\mathcal D _a \mathcal D _b H + h^{(0)}_{ab} H = 0.
\end{equation}
Their bulk generators are given by
\begin{flalign}
	\label{eq:logtransbulkI}
	\xi^\eta & = \log\eta H + \frac{1}{\eta} \log\eta (\mathcal D^aH
	\d_a \sigma - \sigma H) +\frac{2}{\eta} \mathcal D^aH
	\d_a \sigma +  o(\eta^{-1}),\\
	\label{eq:logtransbulkII}
	\xi^a &=\frac{1}{\eta} (\log\eta +1) \mathcal D^a H + \frac{1}{2 \eta^2}
	(\log \eta + \frac{1}{2}) \Big(\mathcal D ^a(\mathcal D ^b H
		\d_b\sigma - H \sigma) + 4 \sigma
	\mathcal D ^aH - k^{ab} \d_b H\Big)\nonumber\\
	& \qquad +\frac{1}{\eta^2}\mathcal D^a\Big( \mathcal D^b H \d_b
\sigma\Big) +
	o(\eta^{-2}).
\end{flalign}
On the hyperboloid, only four linearly independent 
functions satisfy equation \eqref{eq:eqlogtrans}:
\begin{equation}
	H(x^a) = \frac{1}{\sqrt{1-s^2}}\Big(H_0 s + H_1Y^0_{1,-1}(x^A)+
	H_2Y^0_{1,0}(x^A)+ H_3Y^0_{1,1}(x^A)\Big),
\end{equation}
where $Y^0_{lm}$ are usual spherical harmonics. When restricted to these four
functions, logarithmic translations are symmetries of the asymptotic form of
the metric given in \eqref{eq:metricI}. On the other hand, 
transformations of the form \eqref{eq:logtransbulkI}, \eqref{eq:logtransbulkII} 
with an arbitrary function $H(x^a)$ are called logarithmic super-translations.
They don't preserve the asymptotic form of the metric but we can use them to put $\sigma=0$
at the price of the appearance of an extra logarithmic term in $g_{ab}$. These
more general transformations will be useful
in section \ref{sec:scri} when we will make the link with null infinity.

\vspace{5mm} 

The Einstein-Hilbert action, when considered with the set of metrics of the
form \eqref{eq:metricI}, has two major issues: it leads to a divergent
symplectic structure and its variation produces unwanted boundary terms.
Comp\`ere and Dehouck have shown that both problems can be solved by adding
specific boundary terms to the action if one adds the following two conditions
to the asymptotic structure:
\begin{equation}
	\label{eq:boundcondk}
	\mathcal D^a k_{ab} = 0, \qquad k = k^a_{a} = 0.
\end{equation}
These conditions are mild and imposing them does not reduce the set of
solutions under consideration. The first one is the leading term of one of the
bulk equations of motion while the second one can always be reached by a
change of coordinates. It is a form of partial gauge fixing for the SPI
super-translations.

The resulting system has an asymptotic symmetry algebra containing the Lorentz
transformations, a
sub-set of the SPI super-translations and
the logarithmic translations. The remaining SPI super-translations are the ones preserving
\eqref{eq:boundcondk}: they corresponds to the functions $\omega$ satisfying
\begin{equation}
	\label{eq:eqboundsuper}
	(\mathcal D_c \mathcal D^c + 3)	\omega = 0.
\end{equation}
Comp\`ere and Dehouck have computed the associated conserved charges. They are given by
\begin{equation}
	\label{eq:suptranschanges}
	\mathcal Q_\omega = \frac{1}{4\pi G} \oint_{S^2} d^2\Omega \,  (
	\sigma \d_s \omega - \omega \d_s
	\sigma).	
\end{equation}
with the corresponding boundary conserved current
\begin{equation}
	j^a_\omega = \frac{1}{4\pi G} \sqrt{-h^{(0)}} h^{(0)ab} (\omega
	\d_b \sigma - \sigma \d_b \omega).
\end{equation}
The Poincar\'e algebra is a sub-algebra of the 
asymptotic symmetry algebra with the translations forming a sub-algebra of the
remaining super-translations. 
This symmetry structure is very similar to the one we have at
null infinity, however, the super-translations algebra described by
\eqref{eq:eqboundsuper} is bigger than the
BMS4 one and it is a priori not clear how the two are related. 

\vspace{5mm}

The Comp\`ere-Dehouck asymptotic conditions presented here are very relaxed
compared to the ones found in previous works. Their main advantage is that they retain an
infinite dimensional asymptotic symmetry algebra for which we are
still able to define associated conserved charges. For comparison, in their original analysis
of the symplectic structure in \cite{Ashtekar1991}, Ashtekar-Bombelli-Reula introduced the following
extra conditions instead of \eqref{eq:boundcondk}
\begin{equation}
	\label{eq:ABRcond}
	\sigma(-x^a) = \sigma(x^a), \qquad k_{ab} = 0,
\end{equation}
where the first one means that $\sigma$ is even under a combination of time reversal, $s \rightarrow
-s$, and the antipodal map, $x^A \rightarrow -x^A$ or $(\theta, \phi)
\rightarrow (\pi - \theta, \phi + \pi)$. These conditions trivially remove the
divergence of the symplectic structure but they also reduce the asymptotic
symmetries to just Poincar\'e.

\vspace{5mm}

In the following sections, we will always assume that the asymptotic conditions of 
Comp\`ere-Dehouck are satisfied.
Taking this into account, the leading order equations of motion are then
\begin{gather}
	(\mathcal D_c \mathcal D^c + 3)	\sigma = 0, \qquad (\mathcal D_c
	\mathcal D^c - 3)k_{ab} = 0.
\end{gather}
The explicit solutions to these equations can be found in appendix \ref{app:solEDP}.

In order to make the link with null infinity in section \ref{sec:scri}, we will have to
impose extra conditions on the asymptotic fields $\sigma$ and $k_{ab}$
on top of the ones imposed in \eqref{eq:boundcondk}. In particular, we will
have to impose the first condition of Ashtekar-Bombelli-Reula: $\sigma(-x^a) =
\sigma(x^a)$. The main issue is that
some of the solutions do not satisfy the differentiability conditions
usually imposed at null infinity. However, these extra conditions are only needed in order to
make the link with null infinity. The study of the asymptotic symmetry algebra
at spatial infinity presented in the next section does not require them.

\section{BMS4 algebra}
\label{sec:bms4alg}

We have seen in the previous section that, if we forget logarithmic
translations, the asymptotic symmetry
algebra is parametrised by a function $\omega$ and a vector $\mathcal Y^a$ on the hyperboloid
satisfying
\begin{equation}
	(\mathcal D_c \mathcal D^c + 3)	\omega = 0, \qquad L_{\mathcal Y}
	h_{ab}^{(0)} = 0.
\end{equation}
The general solution for $\omega$ is given in the appendix. It takes the form
\begin{gather}
	\omega = \frac{1}{\sqrt{1-s^2}}\left(\hat \omega^{even} +
	\hat\omega^{odd}\right), \\ \hat\omega^{even} = \sum_{l,m}
	\hat\omega^V_{lm}V_l(s)Y^0_{lm}(x^A), \qquad \hat\omega^{odd} =
	\sum_{l,m}\hat\omega^W_{lm} W_l(s)Y^0_{lm}(x^A),
\end{gather}
where $\hat\omega^V_{lm}$ and $\hat\omega^W_{lm}$ are two sets of constants
and the functions $V_l(s)$ and $W_l(s)$ are defined in terms of Legendre
polynomials and Legendre functions of the second kind, respectively.
The functions $\hat \omega^{even}(s,x^A)$ and $\hat \omega^{odd}(s,x^A)$, as
their name indicates, are
even and odd under a combination of time reversal, $s \rightarrow
-s$, and the antipodal map, $x^A \rightarrow -x^A$ or $(\theta, \phi)
\rightarrow (\pi - \theta, \phi + \pi)$. This general solution
is fully characterised by two functions on the 2-sphere:
\begin{flalign}
	T^{even}(x^A) &= \lim_{s\rightarrow 1} \d_s^2 \hat\omega^{even}(s,x^A) = \sum_{l,m}
	\hat\omega^V_{lm}Y^0_{lm}(x^A), \\
	T^{odd}(x^A) &= \lim_{s\rightarrow 1} \hat\omega^{odd}(s,x^A) = \sum_{l,m}
	\hat\omega^W_{lm} Y^0_{lm}(x^A).
\end{flalign}

The vectors $\mathcal Y^a$ represent Lorentz algebra. The rotations are
parametrised by killing vectors of the 2-sphere $\mathcal Y_R^A$ with
$\d_s\mathcal Y_R^A = 0$ and
the corresponding vectors on the hyperboloid are given by
\begin{equation}
	\mathcal Y^s = 0, \quad \mathcal Y^A = \mathcal Y_R^A.
\end{equation}
The boosts are parametrised by functions on the sphere $\psi(x^A)$ such
that $\Delta \psi + 2 \psi = 0$ with their associated vectors on the
hyperboloid being:
\begin{equation}
	\mathcal Y^s = -\frac{1}{2}(1-s^2) \psi, \quad \mathcal Y^A
	=-\frac{1}{2} s \,\gamma^{AB} \d_B
	\psi.
\end{equation}
Rotations and boosts can also be encoded in global conformal killing vectors of the
2-sphere:
\begin{equation}
	Y^A = \mathcal Y^A_R -\frac{1}{2} \gamma^{AB} \d_B \psi, \quad D_A Y^A
	= \psi,
\end{equation}
where $D_A$ is the covariant derivative on the sphere.
This relation forms an isomorphism between the algebra of killing vector fields of the
hyperboloid and the algebra of global conformal killing vector fields of the 2-sphere.

The super-translations combined with the Lorentz algebra form a sub-algebra of the
asymptotic symmetry algebra for which the bracket is given by:
\begin{equation}
	\label{eq:algfullsuplorentz}
	[(\mathcal Y_1, \omega_1),(\mathcal Y_2, \omega_2)] = \Big([\mathcal
	Y_1, \mathcal Y_2], \mathcal Y_1^a \d_a \omega_2 - \mathcal Y_2^a \d_a
	\omega_1\Big),
\end{equation}
or, if we use the rescaled parameter $\hat\omega = \sqrt{1-s^2}\omega$,
\begin{equation}
	[(\mathcal Y_1, \hat\omega_1), 
	(\mathcal Y_2, \hat\omega_2)] =
	\Big([\mathcal
	Y_1, \mathcal Y_2],\mathcal Y_1^a \d_a \hat\omega_2 -
	\frac{s}{2} \psi_1 \hat\omega_2 - (1 \leftrightarrow
	2)\Big).
\end{equation}
As the vector $\mathcal Y^a \d_a$ is even under the combination of a time
reversal and an antipodal map, its action on $\hat\omega$ will not mix the even
and odd parts. Parametrising the super-translations with the two functions on
the sphere $T^{even}(x^A)$ and $T^{odd}(x^A)$ and using the vectors $Y^A$ to
parametrise Lorentz algebra, we can write the bracket as
\begin{equation}
	[(Y_1, T^{even}_1, T^{odd}_1), 
	(Y_2,  T^{even}_2, T^{odd}_2)] = 
	\Big([Y_1,Y_2],\, T^{even}_{[1,2]},\, T^{odd}_{[1,2]}\Big),
\end{equation}
with
\begin{flalign}
	T^{even}_{[1,2]}& = Y_1^A \d_A T^{even}_2 + \frac{3}{2} \psi_1
	T^{even}_2- Y_2^A \d_A T_1^{even} -
	\frac{3}{2} \psi_2 T_1^{even},\\
	\label{eq:actionBMS4}
	T^{odd}_{[1,2]}& = Y_1^A \d_A T^{odd}_2 - \frac{1}{2} \psi_1 T^{odd}_2 - Y_2^A \d_A T_1^{odd} +
	\frac{1}{2} \psi_2 T_1^{odd}.
\end{flalign}

Let's consider the sub-algebra obtained by imposing $T^{even} = 0$. This
algebra is a semi-direct product of an abelian algebra parametrised by an
arbitrary function on the sphere $T^{odd}(x^A)$ with the Lorentz algebra
parametrised by $Y^A$. In \eqref{eq:actionBMS4}, we recognize the action of
Lorentz algebra on the BMS4 super-translations \cite{Barnich2010}. 
This proves that this sub-algebra is isomorphic to
the BMS4 algebra defined at null infinity. We will see in the next section
that, when evaluated at future or past null infinity, the asymptotic killing
vectors associated to $T^{odd}$ reduce to usual BMS4 super-translations.

\vspace{5mm}

We saw in the previous section that the conserved charges associated to super-translations are given by:
\begin{equation}
	\mathcal Q_\omega =\frac{1}{4\pi G} \oint_{S^2} d^2 \Omega(\sigma
	\d_s \omega-\omega \d_s \sigma ).
\end{equation}
Introducing the explicit solution we obtained for $\omega$ and $\sigma$, we
showed in appendix \ref{app:sigom} that these charges take the form:
\begin{equation}
	\mathcal Q_\omega=\frac{1}{8\pi G} \oint_{S^2} d^2\Omega \, \left( T^{odd}m^{even}-T^{even}
	m^{odd} \right),
\end{equation}
where we used the decomposition of $\sigma$
in terms of two functions on the sphere $m^{odd}(x^A)$ and $m^{even}(x^A)$ given in \eqref{eq:defmI}
and \eqref{eq:defmII}.

\section{Null infinity}
\label{sec:scri}

In this section, we will make the connection with null infinity. We will show
that the BMS4 algebra we obtained at spatial infinity is the global BMS4
algebra introduced by Strominger in \cite{Strominger2014}. We will also
recover a linearised version of the results of Herberthson and Ludvigsen relating the 
values of the mass aspect at future and past null infinity
\cite{Herberthson1992}.

\subsection{Structure close to $i_0$}

The description we will use is inspired by the work of 
Friedrich in \cite{Friedrich1998} where he introduced a description
of spatial infinity based on conformal geodesics. The aim was to formulate an initial value
problem for the conformal Einstein equations at spatial-infinity. As 
we will see, it is well adapted to the description of the scattering
problem.

For generic space-times, the asymptotic structure he
obtained close to $i_0$ is as follows
(see \cite{Friedrich1998,Friedrich1999,Friedrich2000} for more details). If some smoothness
conditions on the metric around spatial infinity are satisfied, there exists a 
patch of coordinates $(\rho, s, x^A)$ in a neighbourhood of $i_0$ such that the 
curves obtained by keeping $\rho$ and $x^A$ constant are conformal geodesics.
There exists smooth functions $\tilde \Omega(\rho,x^A)$ and $\omega(\rho,x^A)$ 
such that the rescaled metric $\tilde g_{\mu\nu} = \Omega^2 g_{\mu\nu}$ is
continuous in the limit $s \rightarrow \pm\kappa(\rho)$ where
\begin{equation}
	\label{eq:friedrichconffact}
	\Omega = \tilde\Omega \left(1 -
	\left(\frac{s}{\omega}\right)^2\right),\qquad \lim_{\rho \rightarrow 0}
	\rho^{-1}\tilde \Omega = 1, \quad \lim_{\rho \rightarrow 0} \omega = 1.
\end{equation}
The two hypersurfaces
$s = \pm\omega(\rho, x^A) = \pm 1 + o(\rho^0)$ are then future null
infinity $\mathcal I^+$ and past null
infinity $\mathcal I^-$ while spatial infinity is located at $\rho = 0$.
The rescaled metric
$\tilde\eta_{\mu\nu} = \Omega^2 \eta_{\mu\nu}$ diverges at spatial infinity
but it
is continuous at both null infinities $\mathcal I^\pm$. Slices of
constant time $s$ are spatial hyper-surfaces and in the limit $s$ going to
$\pm\omega$,
these hypersurfaces asymptote to the corresponding $\mathcal I$ (in a neighbourhood of
$i_0$ as these coordinates don't cover the full manifold). Considering the
evolution from a finite time $-\omega<s_0<\omega$ to a finite time
$-\omega<s_1<\omega$ and taking the
limit for infinite times $s_0 \rightarrow -\omega$ and $s_1\rightarrow \omega$, we see
that the in-state and out-state hypersurfaces will naturally contain $\mathcal
I^-$ and $\mathcal I^+$ respectively. A related
feature of these coordinates is that the coordinates at null infinity
are coming from spatial coordinates in the bulk namely $(\rho, x^A)$ while, in
the usual description in terms of the Bondi metric, one of the coordinates is
the asymptote of a time coordinate.
 A few years later in \cite{Friedrich1999,Friedrich2000}, Friedrich
and Kannar made the explicit connection with quantities defined
at null infinity. They for instance computed the Newman-Penrose constants at
null infinity from the coefficients in the expansion around spatial infinity. 

This description relies on a first order formalism of the conformal description of Einstein's
equations. Unfortunately, the link between this formalism and the hyperbolic 
slicing in the metric formalism  of $i_0$ is not simple. In the rest of this
section, we will work in the linearised theory around flat space and we will
describe the background metric using the coordinates obtained by the
Friedrich analysis.

\vspace{5mm}

In order to describe the structure of Minkowski close to $i_0$, we start with the usual hyperbolic
coordinates and introduce the following rescaled radial coordinates $\rho$:
\begin{equation}
	\label{eq:newradialcoord}
	\eta = \frac{1}{\rho \sqrt{1- s^2}}.
\end{equation}
The flat metric then takes the form
\begin{equation}
	\eta_{\mu\nu}dx^\mu dx^\nu = \frac{1}{\rho^2 (1-s^2)^2}
	\left(\frac{1-s^2}{\rho^2} d\rho^2 -2\frac{s}{\rho} d\rho ds -ds^2 +
	\gamma_{AB}dx^A dx^B\right).
\end{equation}
The curves obtained by keeping $\rho$ and $x^A$ constant are conformal
geodesics \cite{Friedrich1995,Friedrich1998}.
The conformal factor \eqref{eq:friedrichconffact} is given in this case by
\begin{equation}
	\label{eq:conffact}
	\Omega = \rho (1-s^2),
\end{equation}
such that the hypersurfaces $\mathcal I^\pm$ are located at $s=\pm 1$. The
metric $q_{ij}$ and the vector field $n^i$ induced on these hypersurfaces by
$\tilde\eta_{\mu\nu} = \Omega^2\eta_{\mu\nu}$ and $\tilde n^\mu = \tilde
\eta^{\mu\nu}\d_\nu\Omega$ are given by
\begin{equation}
	q_{ij} dx^idx^j = \gamma_{AB}dx^Adx^B, \quad n^i \d_i = 2 \rho^2
	\d_\rho,
\end{equation}
where $x^i = (\rho, x^A)$ are the induced coordinates. The usual retarded time
coordinate on $\mathcal I^+$ is given by $u = -\frac{1}{2\rho}$ with $n^i\d_i
= \d_u$.

Let's now consider the metrics introduced in section
\ref{sec:spatialinfinity} and define $h_{\mu\nu} = g_{\mu\nu} -
\eta_{\mu\nu}$ that we will treat as a linear perturbation. Under the change
of coordinates \eqref{eq:newradialcoord}, it takes the form
\begin{multline}
	h_{\mu\nu} = \frac{1}{\rho^2(1-s^2)^2} \left\{ (2\rho  \hat\sigma +
	o(\rho)) \left( \frac{1-s^2}{\rho^2}d\rho^2 - \frac{2s}{\rho}
	dsd\rho\right) + o(\rho) d\rho dx^a \right.\\ \left.+ \rho \sqrt{1-s^2}
	\Big(k_{ab} - 2 \sigma h_{ab}^{(0)}\Big) dx^adx^b + \rho\frac{2s^2}{(1-s^2)}
	\hat\sigma ds^2 + o(\rho) dx^adx^b \right\},
\end{multline}
where $\hat \sigma =\sqrt{1-s^2}\, \sigma$.
Using the explicit solutions obtained  for $\sigma$ and $k_{ab}$ in appendix \ref{app:solEDP}, one can easily check
that $\tilde h_{\mu\nu} = \Omega^2 h_{\mu\nu}$ diverges in the limit
$s\rightarrow \pm 1$. This is related to the asymptotic gauge choice made in
section \ref{sec:spatialinfinity}. To avoid this problem, we will use a different set of coordinates:
\begin{flalign}
	\label{eq:changecoordI}
	\eta &= \eta' - \sigma(x') \log \eta' + o(\eta'^0),\\
	\label{eq:changecoordII}
	x^a & = x'^a - \eta'^{-1}\left(\log\eta'  +
	1\right)(\mathcal D^a \sigma)(x')
	 + o(\eta'^{-1}),
\end{flalign}
such that the metric \eqref{eq:metricI} takes the form:
\begin{gather}
	g'_{\mu\nu} dx'^\mu dx'^\nu = 
	\left( 1 + o(\eta'^{-2})\right)d\eta'^2 + o(\eta'^{-1}) d\eta' dx'^a + 
	g'_{ab}dx'^a dx'^b,\\
	\label{eq:gaugefixedgab}
	g'_{ab} = \eta'^2 h_{ab}^{(0)} +\eta' (\log\eta'+1)\left( -2
	\mathcal D_a \mathcal D_b \sigma - 2\sigma h_{ab}^{(0)}\right) +
	\eta'
	k_{ab}  + o(\eta'),
\end{gather}
where all asymptotic fields are evaluated at $x'^a$. The leading part of
the transformation \eqref{eq:changecoordI}-\eqref{eq:changecoordII} is a 
logarithmic super-translation while the  sub-leading
terms  have to be adapted to reach the asymptotic gauge condition 
chosen here. Super-translations at spatial infinity then take the following form:
\begin{flalign}
	\label{eq:asssymbisI}
	\xi'^{\eta'} &= - \omega + o(\eta'^{-1}), \\
	\label{eq:asssymbisII}
	\xi'^{a}& = -\frac{1}{\eta'} \mathcal D ^a \omega -
	\frac{1}{\eta'^2}(\log \eta' + \frac{3}{2})(\mathcal D ^a
	\mathcal D ^b \sigma + \sigma h^{(0)ab}) \d_b \omega 
	+ \frac{1}{2\eta'^2} k^{ab} \d_b\omega +
	o(\eta'^{-2}).
\end{flalign}
In the rest of this section, we will work with these new
coordinates while removing the primes. 
One remark is in order: while doing this
change of coordinates, we have used elements of the asymptotic symmetry group to
remove a few degrees of freedom. One can check easily that the final metric
\eqref{eq:gaugefixedgab} is independent of $\hat\sigma^W_{00}$ and
$\hat\sigma^W_{1m}$. These modes are the ones on which the logarithmic
translations act. As the transformation generated by
\eqref{eq:changecoordI} and \eqref{eq:changecoordII} is a generalisation of
a logarithmic translation, we can see its action as putting the four modes
$\hat\sigma^W_{l<2, m}$ to zero while transferring all the other modes of
$\sigma$ to the components $g_{ab}$ of the metric. These four modes being
absent, the conjugated super-translations generated by $\hat\omega^V_{l<2,m}$
will have charges equal to zero. In this case, they are proper gauge
transformations.

In these new coordinates, the perturbation $h_{\mu\nu}$ takes the form
\begin{equation}
	\label{eq:metricnull}
	h_{\mu\nu} dx^\mu dx^\nu = \frac{1}{\rho^2
	(1-s^2)^2}\Big\{o(1) d\rho^2
	 + o(\rho)d\rho dx^a + \tilde h_{ab} dx^adx^b\Big\},
\end{equation}
where
\begin{flalign}
	\label{eq:hatgAB}
	\tilde h_{AB} & = -2 \rho (1-s^2)\left(1-\log (\rho\sqrt{1-s^2})
	\right) \left(D_AD_B\hat \sigma + \gamma_{AB} \hat \sigma - s
	\gamma_{AB} \d_s \hat\sigma \right) \nonumber\\ &\qquad \qquad  + \rho
	(1-s^2) \hat k_{AB} + o(\rho),\\
	\tilde h_{As} & = -2 \rho (1-s^2)\left(1-\log (\rho\sqrt{1-s^2})
	\right) \d_A \d_s \hat\sigma+ \rho
	(1-s^2) \hat k_{sA} + o(\rho),\\
	\label{eq:hatgss}
	\tilde h_{ss} & = -2 \rho (1-s^2)\left(1-\log (\rho\sqrt{1-s^2})
	\right) \d_s^2 \hat \sigma  + \rho
	(1-s^2) \hat k_{ss}
	+ o(\rho),
\end{flalign}
with $\hat k_{ab} = \sqrt{1-s^2} \, k_{ab}$. The leading part of the perturbation is now continuous as we approach null infinity and
we have
\begin{equation}
	\lim_{s\rightarrow \pm 1} \tilde h_{ab} = o(\rho).
\end{equation}

In appendix \ref{app:geomunphymet}, we computed the behaviour of the linearised Weyl tensor of 
$\tilde h_{\mu\nu} = \Omega^2 h_{\mu\nu}$ in the limit $s\rightarrow \pm 1$. 
The component that will be relevant for the
definition of the super-momentum charges is given by
\begin{equation}
	\tilde C_{\rho s \rho s} =  - \rho^{-1} (1-s^2) \d_s^2 \hat \sigma +
	o(\rho^{-1}).
\end{equation}
The leading term goes to zero in the limits $s\rightarrow \pm 1$, however, its
rescaled version $K_{\rho s \rho s} = \Omega^{-1}\tilde C_{\rho s \rho s}$ will in general diverge
logarithmically. This divergence breaks the structure of $\mathcal I^\pm$
and we expect that, in these cases, the full metric $g_{\mu\nu}$ does not describe
asymptotically flat space-times at null infinity.
In the following, we will restrict our analysis to 
perturbations $h_{\mu\nu}$ for which the leading term of the rescaled
linearised Weyl tensor $K_{\mu\nu\alpha\beta}$ has a
well defined limit when $s\rightarrow \pm 1$. As shown in the appendix, this restriction imposes $\hat
\sigma_{lm}^W = 0$ for all $l>1$ (it also imposes $\mathcal R
\alpha_{lm}^Q = 0$ for all $l$ where $\mathcal R \alpha_{lm}^Q$ 
characterise part of the
solution of $\hat k_{ab}$). Remark that, as we already put $\hat
\sigma_{00}^W=0=\hat\sigma_{1m}$, we have that the odd part of the function
$\hat\sigma$ is zero:
\begin{equation}
	\hat\sigma^{odd}(s, x^A) = 0 \qquad \Leftrightarrow \qquad m^{odd}(x^A) = 0.
\end{equation}
This implies that the super-translations
parametrised by $T^{even}$ become proper gauge transformations and can be
forgotten. The
asymptotic symmetry algebra at spatial infinity obtained in
section \ref{sec:bms4alg} then reduces to the BMS4 algebra, where the
super-translations are parametrised by $T^{odd}$.

The extra restriction is identical to the first Ashtekar-Bombelli-Reula
extra asymptotic condition given in equation \eqref{eq:ABRcond}. It is also equivalent to the
one made in \cite{Herberthson1992}, where the authors discard one branch of solutions because of
logarithmic divergences in the Weyl tensor when approaching null infinity.

\subsection{Global BMS4 algebra}

At null infinity, super-translations  are vector fields $\xi^+_\alpha$ on $\mathcal I^+$ or $\xi^-_\beta$ on
 $\mathcal I^-$ such that:
\begin{equation}
	(\xi^+_\alpha)^i = \alpha n^i\vert_{\mathcal I^+}, \quad n^i \d_i
	\alpha\vert_{\mathcal I^+} = 0, \qquad
	(\xi^-_\beta)^i = -\beta n^i\vert_{\mathcal I^-}, \quad n^i \d_i
	\beta\vert_{\mathcal I^-} = 0.
\end{equation}
In our case, they are given explicitly by
\begin{equation}
	\xi^+_\alpha = 2\alpha  \rho^2 \d_\rho, \quad
	\xi^-_\beta =-2\beta \rho^2 \d_\rho. 
\end{equation}
Both functions $\alpha$ and $\beta$ are
arbitrary functions on the sphere and are the BMS4
super-translation parameters at future and past null infinity. The
sign we used in the definition of $\xi_\beta^-$ is due to our choice of
coordinates: $\d_\rho$ always points away from spatial infinity.

Around Minkowski and up to a proper gauge transformation, the super-translations 
defined at spatial infinity in \eqref{eq:asssymbisI}-\eqref{eq:asssymbisII}
are given by:
\begin{flalign}
	\xi^\rho & = \rho^2\Big((1+s^2) \hat \omega + s(1-s^2)
	\d_s\hat\omega\Big),\\
	\xi^s & = \rho (1-s^2) \Big((1-s^2) \d_s \hat\omega + s
	\hat\omega\Big),\\
	\xi^A & = -\rho (1-s^2) \gamma^{AB}\d_B\hat\omega.
\end{flalign}
Taking the limit $s\rightarrow \pm1$, we get
\begin{equation}
	\label{eq:limspasuper}
	\lim_{s\rightarrow s^\pm} \xi^\rho = 2\rho^2 \lim_{s\rightarrow \pm 1
	} \hat\omega, \quad 
	\lim_{s\rightarrow s^\pm} \xi^s =0, \quad
	\lim_{s\rightarrow s^\pm} \xi^A =0,
\end{equation}
while, using the form of $\hat \omega$ we obtained in section
\ref{sec:bms4alg} with $T^{even}=0$, we have
\begin{flalign}
	\label{eq:lims1super}
	\lim_{s\rightarrow 1} \hat\omega(s,x^A) & = T^{odd} (x^A),\\
	\label{eq:limsm1super}
	\lim_{s\rightarrow -1} \hat\omega(s,x^A) & = -T^{odd}(-x^A). 
\end{flalign}
These equations show that, on $\mathcal I^+$
and $\mathcal I^-$, the super-translations parametrised by $T^{odd}$ and
originally defined at spatial infinity 
correspond to super-translations
$\xi^+_\alpha$ and $\xi^-_\beta$ defined respectively at future and past null
infinity with
\begin{equation}
	\alpha(x^A) = T^{odd}(x^A),
	\qquad\beta(x^A) = T^{odd}(-x^A).
\end{equation}
This proves that the BMS4 algebra of asymptotic symmetries existing at
spatial infinity when using the asymptotic behaviour described in \eqref{eq:metricI} is
the global BMS4 algebra obtained by A. Strominger in \cite{Strominger2014}. In our
case, the antipodal map between the super-translation parameter at $\mathcal
I^+$ and $\mathcal I^-$ is a consequence of the asymptotics prescribed at
spatial infinity.

\subsection{Super-translation charges}

Other obvious quantities of interest are the associated charges: Bondi
4-momentum and super-translation charges. The component of the linearised Weyl tensor 
relevant for the definition of the super-momentum charges is given by
\begin{equation}
	K_{\rho s \rho s } = \Omega^{-1}\tilde C_{\rho s \rho s} =  - \rho^{-2}\d_s^2 \hat \sigma +
	o(\rho^{-1}).
\end{equation}
At null infinity, we get
\begin{equation}
	\lim_{s\rightarrow \pm 1}K_{\rho s \rho s } = - \rho^{-2}
	\lim_{s\rightarrow \pm 1} \d_s^2 \hat\sigma  + o(\rho^{-2}) = -
	\rho^{-2} m^{even}(\pm x^A) +o(\rho^{-2}),
\end{equation}
where $m^{even}(x^A)$ is a function on the sphere controlling the even
part of $\hat\sigma$. For the Bondi super-momentum charges, we will use the
expression given in \cite{Ashtekar1979}:
\begin{equation}
	P^+_\alpha = \frac{1}{32\pi G}\oint_{S^2} dx^Adx^B \sqrt{-\tilde
	\eta}\epsilon_{\mu\nu\gamma\delta}\,
	K^{\gamma\delta}_{\phantom{\gamma\eta}AB} \alpha \tilde n^\mu
	l^\nu \Big\vert_{\mathcal I^+}, \quad \tilde n^\mu = \tilde
	\eta^{\mu\nu} \d_\nu \Omega,
\end{equation}
where we kept only the terms contributing at the linearised level. The vector
$l^\mu$ is given on $\mathcal I^+$ by $l = \frac{1}{4} \d_\rho +
\frac{1}{2\rho} \d_s$
and $\epsilon_{\alpha\beta\mu\nu}$ is antisymmetric with
$\epsilon_{\rho s \zeta\bar\zeta} = 1$. Evaluating this expression in our case directly leads to:
\begin{flalign}
	P^+_\alpha & =\left. \frac{1}{16\pi G} \oint_{S^2}
	\epsilon_{AB} dx^Adx^B \sqrt\gamma \, \rho \, \alpha \tilde n^\rho\,
	l^s K_{\rho s \rho s}\right\vert_{\mathcal I^+}\\ & = \frac{1}{8\pi G} \oint_{S^2}
	d^2\Omega \Big( \alpha(x^A) m^{even}(x^A) +
	o(\rho^{0})\Big).
\end{flalign}
We are interested in the values of the charges when one approaches spatial
infinity. This corresponds to the limit $\rho \rightarrow 0$. We see that:
\begin{equation}
	\lim_{\rho \rightarrow 0} P^+_\alpha = \frac{1}{8\pi G}  \oint_{S^2}
	d^2\Omega \, \alpha(x^A) m^{even}(x^A).
\end{equation}
These charges are identical to the ones defined at spatial infinity if we take
into account the link between BMS4 super-translation parameters at spatial
and future null infinity: $\alpha(x^A) = T^{odd}(x^A)$.
A similar computation on $\mathcal I^-$ leads to
\begin{gather}
	P^-_\beta = \frac{1}{32 \pi G}\left.\oint_{S^2} dx^Adx^B \sqrt{-\tilde
	\eta}\epsilon_{\mu\nu\gamma\eta}
	K^{\gamma\eta}_{\phantom{\gamma\eta}AB} (-\beta \tilde n^\mu)
	l^\nu\right\vert_{\mathcal I^-},\\ 
	\lim_{\rho \rightarrow 0} P^-_\beta=\frac{1}{8 \pi G}   \oint_{S^2}
	d^2\Omega \, \beta(x^A) m^{even}(-x^A),
\end{gather}
where we used $l\vert_{\mathcal I^-} = - \frac{1}{4} \d_\rho -
\frac{1}{2\rho} \d_s$.
At spatial infinity, the BMS4 super-translations charges defined at future
and past null infinities are equal up to an antipodal map. Explicitly, we
have:
\begin{equation}
	\beta(x^A) = \alpha(-x^A)\qquad \Rightarrow \qquad\lim_{\rho\rightarrow 0} P^+_\alpha =\lim_{\rho\rightarrow 0}
	P^-_\beta.
\end{equation}
This identity is equivalent to the antipodal boundary conditions imposed on the
mass parameter by A. Strominger in \cite{Strominger2014}. We have shown
here that, in the linearised theory, it is a consequence of the boundary conditions imposed at spatial
infinity if we remove the space-times for which the differentiable structure
at null infinity is not strong enough to define the Bondi super-momentum
charges.

This result is a linearised version of a similar result already obtained in 
\cite{Herberthson1992} by Herberthson and Ludvigsen. In their
derivation, they used a generalisation of the conformal description of $i_0$
introduced by Ashtekar and Hansen in \cite{Ashtekar1978}.  It would be
interesting to see how their boundary structure is related to the boundary
condition used in section \ref{sec:spatialinfinity} to
describe spatial infinity.

\section{Conclusions}

In this work, we have shown how a global BMS4 algebra appears as part of the
asymptotic symmetry algebra at spatial infinity. We then used linearised
theory around Minkowski to show that it corresponds to the diagonal algebra considered by
Strominger at null infinity. While obtained in the lagrangian formalism, this is the gravitational equivalent of the
results obtained in \cite{Campiglia2017} for electromagnetism.

The BMS4 charges constructed here are defined on Cauchy slices. It means
that a Hamiltonian description of these charges should also be possible. This would
put this infinite set of conserved charges on the same footing as the ADM
mass.

In section \ref{sec:scri}, we had to rely on linearised theory as the coordinates used to describe 
spatial infinity are not adapted to null infinity. In order to have the full
non-linear picture, it would be of particular interest to rewrite the asymptotic
conditions used in section \ref{sec:spatialinfinity} in the formalism introduced by Friedrich
\cite{Friedrich1995,Friedrich1998}.

In \cite{Barnich2010a,Barnich2010}, it was argued that the relevant asymptotic symmetry algebra at null
infinity should not only contain Lorentz algebra but the full conformal algebra 
on the 2-sphere. In that case, it has been shown that the relevant structure
is an algebroid and that the associated algebroid of charges closes up to a
central extension \cite{Barnich2011,Barnich2013,Barnich2017}. It would be
interesting to see if one can reproduce this structure at spatial infinity. An
interesting result in this direction has been given in
\cite{Baghchesaraei2016} where the authors present a set of boundary
conditions at spatial infinity with an extended form of BMS4 as asymptotic
symmetry algebra. As this algebra does not contain boosts nor spatial
translations, it does not describe the same transformations as the ones
obtained in this work but it may provide hints for possible generalisations.

\section*{Acknowledgements}

I would like to thank G. Barnich, J. Korovins and T. Lessinnes for useful
discussions. 

\appendix

\section{Solution to some differential equations}
\label{app:solEDP}

In this appendix, we will solve the various partial differential equations
relevant for our asymptotic analysis. The equation of motion for $\sigma$ as
well as the equation satisfied by super-translation parameter are 
\begin{equation}
	\label{eq:appeqsigom}
	(\mathcal D_a
	\mathcal D^a + 3) \sigma = 0, \qquad (\mathcal D_a
	\mathcal D^a + 3) \omega = 0,
\end{equation}
while the asymptotic field $k_{ab}$ satisfies
\begin{equation}
	\label{eq:appkresume}
	k^a_{\phantom aa} = 0, \quad \mathcal D^a k_{ab} = 0, \quad
	(\mathcal D_a \mathcal D^a - 3)k_{bc} = 0.
\end{equation}
As in section \ref{sec:bms4alg} and \ref{sec:scri}, we will use the rescaled quantities:
\begin{equation}
	\hat \sigma = \sqrt{1-s^2}\, \sigma, \quad 
	\hat \omega = \sqrt{1-s^2}\, \omega, \quad 
	\hat k_{ab} = \sqrt{1-s^2}\, k_{ab}.
\end{equation}

In order to solve these equations, we will use complex coordinates on the
sphere $\zeta = \cot \frac{\theta}{2} e^{i\phi}$ for which the metric takes
the form
\begin{equation}
	\label{eq:appeqk}
	\gamma_{AB} dx^Adx^B = 2 P^{-2} d\zeta d\xbar\zeta, \qquad P =
	\frac{1+\zeta\xbar\zeta}{\sqrt 2}.
\end{equation}
Tensors on the sphere can be encoded in spin weighted functions $\eta$ of spin
$\mathbcal s_\eta$ and the covariant
derivative is then given by the operators
\begin{equation}
	\eth \eta = P^{1-\mathbcal s_\eta} \d_{\bar \zeta} (P^{\mathbcal s_\eta} \eta), \qquad
	\xbar\eth \eta = P^{1+\mathbcal s_\eta} \d_\zeta(P^{-\mathbcal s_\eta} \eta),
\end{equation}
where $\eth, \xbar\eth$ respectively raises and lowers the spin weight by
one unit (see \cite{Penrose1984,Penrose1986} for more details). They satisfy
\begin{equation}
	[\xbar \eth, \eth] \eta = s_\eta \eta,
\end{equation}
and the Laplace operator on the sphere can be written as
\begin{equation}
	\Delta = \xbar\eth \eth + \eth\xbar\eth.
\end{equation}
The asymptotic fields $\hat\sigma$ and $\hat\omega$ are spin weighted
functions of spin zero while the tensor $\hat k_{ab}$ can be
encoded in the following spin weighted functions:
\begin{gather}
	\hat k_{ss} = \kappa, \quad \hat k_{s\zeta} = P^{-1}
	\alpha, \quad  \hat k_{s\bar\zeta} = P^{-1}
	\xbar\alpha, \\ 
	 \hat k_{\zeta\zeta} = P^{-2}
	\beta, \quad
	 \hat k_{\bar\zeta\bar\zeta} = P^{-2}
	\xbar\beta, \quad
	\hat k_{\zeta\bar\zeta} = \frac{P^{-2}}{2}
	(1-s^2) \kappa, 
\end{gather}
where we used the first equation of \eqref{eq:appkresume}. The spin weights
are given by
\begin{equation}
	 \mathbcal s_{\kappa} = 0,
	\quad   \mathbcal s_\alpha = -1,\quad  \mathbcal s_{\bar\alpha} = 1,
	\quad \mathbcal s_{\beta} = -2,
	\quad \mathbcal s_{\bar \beta} = 2.
\end{equation}
The functions $\kappa$ and $\sigma$ are real while $\alpha, \xbar \alpha, \beta$ and
$\xbar\beta$ satisfy $\alpha ^* = \xbar \alpha$ and $\beta ^* = \xbar \beta$
where the star denotes complex conjugation.

The two equations in \eqref{eq:appeqsigom} can then be written as
\begin{equation}
	\label{eq:appeqsighat}
	-(1-s^2) \d_s^2 \hat \sigma - 2s\d_s \hat \sigma + 2\hat \sigma  +
	(\eth\xbar\eth + \xbar\eth \eth) \hat\sigma=
	0,
\end{equation}
with the same equation for $\hat \omega$. The various equations for $k_{ab}$
take the form
\begin{flalign}
	\label{eq:appdivs}
	(1-s^2) \d_s \kappa & = \eth \alpha + \xbar \eth \xbar \alpha, \\
	\label{eq:appdivzeta}
	(1-s^2) \d_s \alpha + s \alpha & = \frac{1}{2} (1-s^2)\xbar\eth \kappa + \eth
	\beta,
\end{flalign}
associated with
\begin{flalign}
	\label{eq:appeomss}
	-(1-s^2) \d_s^2 \kappa + 2s \d_s \kappa + (\eth \xbar \eth + \xbar
	\eth \eth) \kappa & = 0, \\
	\label{eq:appeomsz}
	-(1-s^2) \d_s^2 \alpha + 2s \d_s \alpha - \alpha + (\eth \xbar \eth + \xbar
	\eth \eth) \alpha & =  2 s \xbar \eth \kappa, \\
	\label{eq:appeomzz}
	-(1-s^2) \d_s^2 \beta + 2s \d_s \beta - 2 \beta + (\eth \xbar \eth + \xbar
	\eth \eth) \beta & = 4 s \xbar \eth \alpha.
\end{flalign}
To this set, we have to add the equivalent of equations \eqref{eq:appdivzeta}, \eqref{eq:appeomsz} and \eqref{eq:appeomzz} for the barred quantities.

In order to solve these equations, we will expend our spin weighted functions in spin
weighted spherical harmonics $Y^s_{lm}(x^A)$ where $Y^0_{lm}(x^A)$ are the
usual spherical harmonics. Spin weighted spherical harmonics are only defined
for $l> \vert s \vert$ and $l > \vert m \vert$. They
form a complete set for each value of the spin $s$. The main properties we will be
using are:
\begin{gather}
	\eth\, Y^s_{lm} = - \sqrt{\frac{(l-s)(l+s+1)}{2}}\,  Y^{s+1}_{lm}, \quad
	\xbar\eth\, Y^s_{lm} = \sqrt{\frac{(l+s)(l-s+1)}{2}}\, Y^{s-1}_{lm}, \\
	(\eth\xbar\eth + \xbar\eth \eth)\, Y^s_{lm} = -[(l+1)l - s^2]
	\,Y^s_{lm}, \quad (Y^s_{lm})^* = (-1)^{m+s}Y^{-s}_{l,-m}.
\end{gather}

\subsection{Solution for $\sigma$ and $\omega$}
\label{app:sigom}

We will focus on the solution to the equation of motion for $\sigma$ as the
equation satisfied by the super-translation parameter $\omega$ is identical.

Introducing the spherical harmonic decomposition $\hat \sigma = \sum_{l,m} \hat\sigma_{lm}(s)\, Y^0_{lm}$, equation
\eqref{eq:appeqsighat} becomes 
\begin{equation}
	\label{app:eqorigpsi}
	-(1-s^2) \d_s^2 \hat \sigma_{lm} - 2s\d_s \hat \sigma_{lm} + 2\hat
	\sigma_{lm}  -l (l+1) \hat\sigma_{lm}=
	0, \quad \forall l,m.
\end{equation}
This equation is related to Legendre equation:
\begin{equation}
	\label{app:eqlegendre}
	-(1-s^2) \d_s^2 \psi_l + 2s\d_s \psi_l - l(l+1)\psi_l = 0
\end{equation}
for which the general solution is given in terms of Legendre polynomials
$P_l(s)$ and Legendre functions of the second kind $Q_l(s)$:
\begin{gather}
	\psi_l(s) = \psi_l^P P_l(s) + \psi_l^Q Q_l(s),\\
	\label{eq:legendre2kind}
	Q_l(s) = P_l(s) \frac{1}{2} \log \left(\frac{1+s}{1-s}\right) + \tilde
	Q_l(s),
\end{gather}
where $\tilde Q_l$ are polynomials (see \cite{Szego1991}).
One can easily show that if $\psi_l$ satisfies \eqref{app:eqlegendre} then
$(1-s^2)\d_s^2 \psi_l$ 
will satisfy  equation \eqref{app:eqorigpsi}. If we define
\begin{equation}
	V_l(s) =(l-1)l(l+1)(l+2)  (1-s^2)^2 \d_s^2 P_l,\quad W_l(s) = \frac{1}{2}
	(1-s^2)^2 \d_s^2 Q_l, \qquad \forall l>1,
\end{equation}
the general solution to \eqref{app:eqorigpsi} is then given by
\begin{equation}
	\label{eq:solorigpsi}
	\hat\sigma_{lm}(s) = \hat\sigma^V_{lm}V_l(s) + \hat\sigma^W_{lm}
	W_l(s), \qquad \forall l>1.
\end{equation}
The normalisations of $V_l$ and $W_l$ were chosen for future convenience. For $l<2$,
this procedure gives us only one of the two independent solutions namely:
\begin{equation}
	W_0(s) = s, \qquad W_1(s) = 1.
\end{equation}
The other one is easily constructed:
\begin{equation}
	V_0= \frac{1}{2}(s^2 +1), \qquad V_1 = \frac{1}{6}(s^3 - 3s).
\end{equation}
The general solution to the equation of motion for $\hat\sigma$ then takes the form
\begin{equation}
	\label{eq:gensolsig}
	\hat\sigma(s,x^A) = \sum_{l,m} \left(\hat\sigma^V_{lm}V_l(s) +
	\hat\sigma^W_{lm}
	W_l(s)\right) Y^0_{lm}(x^A).
\end{equation}

The functions $V_l$ and $W_l$ inherit the parity properties of $P_l$ and
$Q_l$:
\begin{flalign}
	P_l(-s) = (-1)^l P_l(s) \qquad &\Rightarrow \qquad V_l(-s) = (-1)^l V_l(s),\\
	Q_l(-s) = - (-1)^l Q_l(s) \qquad &\Rightarrow \qquad W_l(-s) = -(-1)^l W_l(s).
\end{flalign}
This means that under the combined action of a time reversal $s\rightarrow -s$ and an
antipodal map $x^A\rightarrow -x^A$, the general solution \eqref{eq:gensolsig} separates into an odd and
an even part:
\begin{gather}
	\hat \sigma = \hat\sigma^{even} +
	\hat\sigma^{odd}, \quad \hat \sigma^{even}(-s,-x^A) =
	\hat\sigma^{even}, \quad \hat \sigma^{odd}(-s,-x^A) =
	-\hat\sigma^{odd},\\
	\hat\sigma^{even} =\sum_{l,m}\hat\sigma^V_{lm}V_l(s) Y^0_{lm}(x^A), 
	\quad \hat\sigma^{odd} = \sum_{l,m}
	\hat\sigma^W_{lm}
	W_l(s) Y^0_{lm}(x^A).
\end{gather}
Each these parts can be parametrised by a function on the sphere:
\begin{flalign}
	\label{eq:defmI}
	m^{even}(x^A) & \equiv \lim_{s\rightarrow
	1}\sum_{l,m}\hat\sigma^V_{lm}\d_s^2V_l(s) Y^0_{lm}(x^A)=
	\sum_{l,m}\hat\sigma^V_{lm}Y^0_{lm}(x^A), \\
	\label{eq:defmII}
	m^{odd}(x^A) & \equiv \lim_{s\rightarrow
	1}\sum_{l,m}\hat\sigma^W_{lm}W_l(s) Y^0_{lm}(x^A)=
	\sum_{l,m}\hat\sigma^W_{lm}Y^0_{lm}(x^A), 
\end{flalign}
where we used the following identities 
\begin{equation}
	\lim_{s\rightarrow 1} \d_s^2 V_l = 1, \qquad \lim_{s\rightarrow 1} W_l
	= 1.
\end{equation}
They can be easily shown using $P_l(1) = 1$, the explicit form of $Q_l$ given in
\eqref{eq:legendre2kind} and Legendre equation \eqref{app:eqlegendre}.
Doing an asymptotic expansion around $s=1$ of both parts of the solution, we
get
\begin{flalign}
	\hat \sigma^{odd} (s,x^A) &= m^{odd}(x^A) + O\Big((1-s)\Big),\\
	\hat \sigma^{even} (s,x^A) &= \hat \sigma^V_{00} -\frac{1}{3} 
	\sum^{m=1}_{m=-1} \hat\sigma_{1m}^V
	Y^0_{1m}(x^A) + O\Big((1-s)\Big).
\end{flalign}
A similar expansion can be done around $s=-1$.

We will have the same kind of expressions for the super-translation parameter: $\hat\omega = \hat\omega^{even} +
\hat\omega^{odd}$ with
\begin{gather}
	\hat\omega^{even} =\sum_{l,m}\hat\omega^V_{lm}V_l(s) Y^0_{lm}(x^A), 
	\quad \hat\omega^{odd} = \sum_{l,m}
	\hat\omega^W_{lm}
	W_l(s) Y^0_{lm}(x^A),\\
	\label{app:defT}
	T^{even}(x^A) \equiv
	\sum_{l,m}\hat\omega^V_{lm}Y^0_{lm}(x^A), \quad
	T^{odd}(x^A) \equiv
	\sum_{l,m}\hat\omega^W_{lm}Y^0_{lm}(x^A).
\end{gather}

Super-translation charges given in \eqref{eq:suptranschanges} can be rewritten as
\begin{equation}
	\mathcal Q_\omega = \frac{1}{4\pi G} \oint_{S^2} d^2\Omega \,
	\frac{1}{1-s^2}(\hat\sigma \d_s \hat\omega - \hat\omega \d_s
	\hat\sigma).	
\end{equation}
Inserting the general solutions we obtained, we get
\begin{flalign}
	\mathcal Q_\omega &= \frac{1}{4\pi G} \sum_{lm}\,
	\left(\xbar{\hat\omega_{lm}^W}\hat\sigma^V_{lm} -
	\xbar{\hat\omega_{lm}^V} \hat\sigma^W_{lm}\right)\mathcal C_l,\\
	\mathcal C_l & = \frac{1}{1-s^2}
	\left(V_l \d_s W_l - W_l \d_sV_l\right).
\end{flalign}
The quantity $\mathcal C_l$ is conserved $\d_s \mathcal C_l = 0$ and its value is
easily computed asymptotically: $\mathcal C_l = \lim_{s\rightarrow 1} \mathcal
C_l = \frac{1}{2}$. Plugging this into the value of the charges and using the
functions introduced in \eqref{eq:defmI},\eqref{eq:defmII} and \eqref{app:defT}, we get
\begin{equation}
	\mathcal Q_\omega = \frac{1}{8\pi G} \oint_{S^2} d^2 \Omega \,
	T^{odd}(x^A) m^{even}(x^A) - \Big(T^{even}(x^A) m^{odd}(x^A)\Big).
\end{equation}

\subsection{Solution for $k_{ab}$}
\label{app:solk}

Let's now have a look at the equations of $k_{ab}$: equations \eqref{eq:appdivs} to
\eqref{eq:appeomzz}. We will introduce the corresponding spherical harmonic
expansions:
\begin{gather}
	\kappa = \frac{1}{1-s^2}\hat k = \sum_{l,m} \kappa_{lm} Y^0_{lm},
	\quad \alpha =
	\sum_{l>0,m} \alpha_{lm} Y^{-1}_{lm}, \quad \xbar\alpha =
	\sum_{l>0,m} \xbar\alpha_{lm}Y^{1}_{lm},\\ \beta = \sum_{l>1,m} \beta_{lm}
	Y^{-2}_{lm}, \qquad \xbar\beta = \sum_{l>1,m}
	\xbar\beta_{l,m}
	Y^{2}_{lm},
\end{gather}
where the reality conditions imply 
\begin{equation}
	(\kappa_{lm})* = (-1)^m
	\kappa_{l,-m},\quad (\alpha_{lm})^* = -(-1)^m \xbar\alpha_{l,-m},\quad 
	(\beta_{lm})^* = (-1)^m \xbar\beta_{l,-m}.
\end{equation}
Inserting this into equations \eqref{eq:appeomss} and \eqref{eq:appeomsz}, we
get
\begin{flalign}
	\label{eq:eqmodkappa}
	-(1-s^2) \d_s^2 \kappa_{lm} + 2s \d_s \kappa_{lm} - l(l+1) \kappa_{lm} & =
	0, \qquad \qquad \qquad \qquad \forall l,m,\\
	\label{eq:eqmodalpha}
	-(1-s^2) \d_s^2 \alpha_{lm} + 2s \d_s \alpha_{lm} - l(l+1) \alpha_{lm} & =
	 s \sqrt{2l(l+1)} \kappa_{lm}, \quad \forall l>0,m,\\
	\label{eq:eqmodbaralpha}
	-(1-s^2) \d_s^2 \xbar\alpha_{lm} + 2s \d_s \xbar\alpha_{lm} - l(l+1)
	\xbar\alpha_{lm} & =
	-s \sqrt{2l(l+1)} \kappa_{lm}, \quad \forall l>0,m,
\end{flalign}
while equation \eqref{eq:appdivs} gives
\begin{equation}
	\label{eq:eqmodkappaalpha}
	(1-s^2) \d_s \kappa_{00} = 0, \quad (1-s^2) \d_s \kappa_{lm} = \sqrt{\frac{(l+1)l}{2}}( \xbar\alpha_{lm}- 
	\alpha_{lm}), \quad \forall l>0,m.
\end{equation}
Both $\kappa_{lm}$ and $\alpha_{lm} + \xbar\alpha_{lm}$ satisfy Legendre
equation for which the general solution is given in terms of Legendre polynomials
$P_l$ and Legendre functions of the second kind $Q_l$:
\begin{equation}
	Q_l(s) = P_l(s) \frac{1}{2} \log \left(\frac{1+s}{1-s}\right) + \tilde
	Q_l(s),
\end{equation}
where $\tilde Q_l$ are polynomials. The general solution to equations
\eqref{eq:eqmodkappa}, \eqref{eq:eqmodalpha}
and \eqref{eq:eqmodkappaalpha} is then given by:
\begin{gather}
	\label{eq:appsolI}
	\kappa_{00}(s) = \kappa^P_{00}, \quad 
	\kappa_{lm}(s) = \kappa^P_{lm} P_l(s) + \kappa^Q_{lm} Q_l(s) \qquad
	\forall l>0,m, \\
	\label{eq:appsolII}
	\mathcal R\alpha_{lm}(s) = \frac{1}{2}\left(\alpha_{lm}(s) + \xbar
	\alpha_{lm}(s)\right) = \mathcal  R\alpha^P_{lm} P_l(s) +
	\mathcal R\alpha^Q_{lm} Q_l(s) \quad
	\forall l>0,m, \\
	\label{eq:appsolIII}
	\alpha_{lm}(s) =\frac{-1}{\sqrt{2l(l+1)}} (1-s^2) \d_s\kappa_{lm} + \mathcal
	R\alpha_{lm}, \quad \forall l>0, m.
\end{gather}
Developing equation \eqref{eq:appdivzeta}, we then get
\begin{gather}
	(1-s^2)\d_s \alpha_{lm} + s \alpha_{lm} - \frac{1}{2}
	\sqrt{\frac{l(l+1)}{2}} (1-s^2) \kappa_{lm} = - 
	\sqrt{\frac{(l+2)(l-1)}{2}} \beta_{lm}, \quad \forall l>1,m,\\
	(1-s^2) \d_s \alpha_{1m} + s \alpha_{1m} - \frac{1}{2}
	(1-s^2) \kappa_{1m} = 0.  
\end{gather}
which, when associated to their barred equivalent, lead to
\begin{gather}
	\label{eq:appsolIV}
	\kappa^Q_{1m} = 0, \quad \mathcal R \alpha_{1m}^P = 0, \quad \mathcal
	R \alpha_{1m}^Q = 0, \\
	\label{eq:appsolV}
	\beta_{lm}(s) = \sqrt{\frac{2}{(l-1)(l+2)}} \left[
		\frac{1}{2}\frac{(1-s^2)^2}{\sqrt{2l(l+1)}} \d_s^2 \kappa_{lm}
		- \left((1-s^2) \d_s + s \right) \mathcal R\alpha_{lm}\right],
\end{gather}
where the last line is valid for $l>1$. One can check that equation
\eqref{eq:appeomzz} is then automatically satisfied. Equations
\eqref{eq:appsolI}-\eqref{eq:appsolIII} with equations \eqref{eq:appsolIV} and
\eqref{eq:appsolV} give the complete solution to the system of equations
\eqref{eq:appdivs} to \eqref{eq:appeomzz}. 

\section{Weyl tensor of the unphysical metric}
\label{app:geomunphymet}

This appendix contains various useful results about geometric quantities
associated to the unphysical metric $\tilde g_{\mu\nu} = \tilde \eta_{\mu\nu}
+ \tilde h_{\mu\nu}$ written in equations
\eqref{eq:metricnull} to \eqref{eq:hatgss}:
\begin{equation}
	h_{\mu\nu} dx^\mu dx^\nu = \frac{1}{\rho^2
	(1-s^2)^2}\Big\{o(1) d\rho^2
	 + o(\rho)d\rho dx^a + \hat h_{ab} dx^adx^b\Big\},
\end{equation}
where
\begin{flalign}
	\tilde h_{AB} & = -2 \rho (1-s^2)\left(1-\log (\rho\sqrt{1-s^2})
	\right) \left(D_AD_B\hat \sigma + \gamma_{AB} \hat \sigma - s
	\gamma_{AB} \d_s \hat\sigma \right) \nonumber\\ &\qquad \qquad  + \rho
	(1-s^2) \hat k_{AB} + o(\rho),\\
	\tilde h_{As} & = -2 \rho (1-s^2)\left(1-\log (\rho\sqrt{1-s^2})
	\right) \d_A \d_s \hat\sigma+ \rho
	(1-s^2) \hat k_{sA} + o(\rho),\\
	\tilde h_{ss} & = -2 \rho (1-s^2)\left(1-\log (\rho\sqrt{1-s^2})
	\right) \d_s^2 \hat \sigma  + \rho
	(1-s^2) \hat k_{ss}
	+ o(\rho).
\end{flalign}
The linearised Weyl tensor of $\tilde h_{\mu\nu}$ is given by
\begin{flalign}
	\tilde C_{\rho a \rho b} &= -\frac{1}{\rho} (1-s^2) \hat\sigma_{ab} +
	o(\rho^{-1}),\\
	\tilde C_{\rho a b c} & = (1-s^2) \left( \frac{3}{2}
	\sqrt{1-s^2} (\mathcal D_b k_{ac} - \mathcal D_c k_{ab}) +
	\frac{s}{1-s^2} (\delta^s_b \hat \sigma_{ac} - \delta^s_c \hat
	\sigma_{ab})\right) + o(1).
\end{flalign}
All the other components can be obtained using the properties of the Weyl
tensor. The combination relevant for the description of null infinity is
$K_{\mu\nu\alpha\beta}=\Omega^{-1} \tilde C_{\mu\nu\alpha\beta}$. If this tensor
is not continuous at null infinity then the structure of $\mathcal I^\pm$ is
not differentiable enough to allow the definition of the BMS4 super-translation
charges. Let's have a look at a few specific components:
\begin{flalign}
	K_{\rho s \rho s }& = -\frac{1}{\rho^2} \d_s^2\hat\sigma +
	o(\rho^{-2}),\\
	K_{\rho s \zeta\bar\zeta} & = \frac{1}{\rho} \frac{3}{2} P^{-2} ( \xbar
	\eth \xbar \alpha - \eth \alpha) + o(\rho^{-1}) \\
	& = - \frac{3}{\rho}
	 P^{-2} \sum_{lm} \sqrt{\frac{l(l+1)}{2}} \left(\mathcal R \alpha^P_{lm} P_l(s)
	 +
	 \mathcal R \alpha^Q_{lm}Q_l(s)\right)Y^0_{lm}+ o(\rho^{-1}).
\end{flalign}
In the limit $s \rightarrow \pm 1$, these components diverge logarithmically when
$\hat\sigma^W_{lm} \ne 0$ and $\mathcal R \alpha_{lm}^Q \ne 0$.

\bibliography{./biblio}

\end{document}